\documentclass[twocolumn,showpacs,preprintnumbers,amsmath,amssymb]{revtex4}

\usepackage{graphicx}
\usepackage{dcolumn}
\usepackage{bm}

\begin{document}

\title{The importance of local band effects for ferromagnetism 
in hole doped La$_2$CuO$_4$.}

\author{B. Barbiellini$^1$ and T. Jarlborg$^2$}

\affiliation{$1$ Department of Physics, 
Northeastern University, Boston, 
Massachusetts 02115, USA \\
$2$ DPMC, University of Geneva, 24 Quai Ernest-Ansermet, 
CH-1211 Geneva 4, Switzerland}


\begin{abstract} 

Band calculations for supercells of La$_{(2-x)}$Ba$_x$CuO$_4$ 
show that the rigid band model for doping is less adequate than what is 
commonly assumed. In particular, weak ferromagnetism (FM) can appear locally 
around clusters of high Ba concentration. The clustering is important at 
large dilution and averaged models for magnetism, such as the virtual 
crystal approximation, are unable to stabilize magnetic moments. These 
results give a support to the idea that weak FM can be the cause of the 
destruction of superconductivity at high hole doping.

\end{abstract}


\maketitle

The fascinating phase diagram of the cuprate high temperature 
superconductors still remains a great mystery in the field of 
condensed-matter physics \cite{cho}. 
Antiferromagnetic (AF) spin fluctuations
are probably also relevant to the properties of 
the doped metallic state because they 
mix with phonons \cite{tjrap,tjb} 
and they provide 
mechanisms able to boost the electron-phonon interaction 
\cite{gunnarsson08}. 
While many investigations 
have focused on underdoped phases 
\cite{carlson,cvekovic}, little attention 
has been devoted to the overdoped 
regime, where the abrupt suppression 
of superconductivity for a critical 
doping cannot be easily understood 
in term of the standard BCS theory. 
Only recently, Kopp {\em et al.} \cite{kopp} have suggested that ferromagnetic 
(FM) fluctuations compete with superconductivity in the overdoped regime. 
These authors have considered an itinerant picture and have used it to 
justify the rich phenomenology of the overdoped cuprates. In particular, 
the termination of the superconductivity in the overdoped regime could 
mark a quantum critical point beyond which there should be a FM phase at 
zero temperature.
A FM groundstate excludes a $s$ or $d$ wave superconductivity of the 
singlet variety \cite{kopp}.  The electron-phonon coupling, which may 
contribute to superconductivity, is also inhibited by FM spin fluctuations 
\cite{daams,tj2003}. 

Electronic structure calculations performed by Perry {\em et 
al.}  \cite{per0} using the spin unrestricted 
Becke-3-Lee-Yang-Parr (B3LYP) 
hybrid functional successfully produced a 2.0 eV band gap and AF 
in (undoped) La$_2$CuO$_4$. 
However, Paier 
{\em et al.} \cite{paier} have raised serious concerns on the suitability 
of the B3LYP potential in metallic phases, and
subsequent B3LYP calculations for minimal size supercells of
La$_{2-x}$Sr$_x$CuO$_4$, corresponding to $x$ = 0.125, 0.25, 
and 0.5, \cite{per} have predicted FM solutions. This in contrast to the
non-magnetic or AF fluctuating states found in real, weakly doped systems.
On the other hand, band calculations based on 
the local spin-density approximation (LSDA) 
yields the correct Fermi surface and a reasonable band dispersion 
\cite{seppo}, and electronic 
structure studies on manganites \cite{mijna} have shown that the  
LSDA provides reliable potential for 
predicting both metallic and ferromagnetic phases. 
The difficulty with
LSDA is rather to describe correctly the AF, insulating state of the undoped
cuprates, but one should note that LSDA calculations for high-T$_c$ cuprates with
{\em off-center} linearization energies describe well both the undoped 
and the doped counterparts \cite{jpcm}. This fact indicates that two very 
different ground states 
are almost degenerate and that only small 
LSDA-corrections are needed to 
bring the two states into the correct order \cite{pickett}.

In this letter, we confirm that 
the LSDA electronic structure 
of La$_{2-x}$Ba$_x$CuO$_4$ 
in the overdoped 
regime is consistent with weak ferromagnetism 
appearing locally around clusters of high Ba concentration.
Our results from first principles validate the 
hypothesis by Kopp {\em et al.} \cite{kopp}
that FM order and superconductivity are competing
in overdoped samples


Our band calculations are performed with the Linear Muffin-Tin Orbitial 
(LMTO) method \cite{lmto,ja,bdj} for supercells with two or 16 formula units 
(f.u.) of La$_2$CuO$_4$. The elementary 7-site unit cell is doubled along 
the $\hat{x}$ direction in the former case, which yields a (2,1,1)-extension of the 
elementary cell. The cell is then doubled along the 3 directions in the 
case of the large cell, leading to an (4,2,2)-extended cell
with 112 atomic sites totally.
The supercells in Ref.~\cite{per} were adjusted to optimal size 
for each doping $x$, while here the use of a larger supercell of 
identical dimension for different $x$ 
allow for some randomness in the local order.
The basis set includes $\ell$ up through 3 for La- and Cu-sites, and up 
through 2 for O-sites \cite{footnote1}.  
The number of 
$k$-points is 84 in the calculations for the small cell and it varies from 4 
to 125 in 1/4 of the Brillouin zone for the large cell.
All sites in the cell are considered as non-equivalent 
throughout the self-consistent calculations. Other details of 
the methods can be found elsewhere \cite{lmto,ja,bdj,san,spc}.

Our result for paramagnetic La$_2$CuO$_4$ is in excellent agreement with
the full potential calculation by Yu {\em et al.} \cite{yu}.
As was mentioned above, doped La$_2$CuO$_4$ is metallic and 
LSDA calculations describe its band dispersion
and Fermi surface adequately,  and AF can be promoted by 
use of off-set linearization
energies in LMTO. The present calculations are based 
on a normal choice of linearization energies, and
therefore we expect an underestimation of the tendency towards magnetism, 
even in the doped cases.

%
%
%
The magnetic moments for the three possible configurations for $x=1$ in
the supercells with two f.u.(14-site cells)
are shown in Table \ref{table1}. Small but 
stable magnetic moments develop in the two configurations 
where the two Ba atoms in the cell belong to 
the same plane (i.e. are situated next to each other along $\hat{x}$ and 
$\hat{y}$), or when they are shared between two planes. The two Cu in both configurations
have almost equal magnetic moments and equal local 
density-of-states (DOS) at the Fermi energy $E_F$. 
Planar oxygen atoms have small magnetic moments, but the apical 
oxygen atoms between La/Ba and Cu has almost 
as large magnetic moment as Cu. Oxygen magnetism is 
somewhat unusual, but there are cases reported in the literature
like for a half-metallic Rb$_4$O$_6$ \cite{attema}. 
This participation of the apical  O 
to the spontaneous magnetic polarization cannot be understood 
from the Stoner model because there is no clear correlation between 
local DOS at the apical O and the magnetic moment.  
Therefore, other mechanisms such as charge transfers 
and hybridization with neighbors are vital to explain 
the magnetization of the apical O atom \cite{per}. 
In our case, the charges
on the apical O atoms are somewhat larger when these O sites are between 
the Cu atom and the Ba atom rather than between Cu atom and La atom, 
while the magnetic moments are largest on the O atom 
close to the La-plane when the La/Ba atoms are fully occupying a plane.
This trend is reversed when there
is alternate La/Ba occupation within the planes.
No magnetic moment is developed when the 
two Ba atoms are located on top of each other, i.e. along $\hat{z}$. This 
last result might seem surprising, since a finite magnetic moment is present next 
to a single Ba atom in the two other cases. 

Therefore, it seems that
inhomogeneous formation of doped holes in the vicinity of the impurity and
the clustering of impurity atoms within the same plane are important 
ingredients for the formation of magnetic moments on 
Cu atoms and apical O sites. 
This conclusion is corroborated by
the fact that no magnetism is found when the virtual crystal approximation (VCA) 
is applied for $x=1$. The missing charge on Ba sites relative to La sites 
is only partly taken from the Cu band, so $E_F$ does not quite reach 
the high DOS region, thus the system is below the Stoner limit for FM. 
Eventually, when the Ba concentration becomes larger ($x \geq 1.5$), FM appears
in VCA calculations.
Random oxygen vacancies, described by CPA calculations, will increase 
the DOS at $E_F$ \cite{papaconstantopoulos}, and might help Stoner magnetism.

Next, by using larger supercell calculations with Ba/La substitutions, 
we try to answer the question if local spontaneous magnetization is possible 
for hole doping corresponding to the most "over-doped" (actually
meaning low doping for supercell calculations) high-T$_c$ materials, 
and thus if FM can be responsible for the disappearance of superconductivity.
\begin{table}[b]
\caption{\label{table1} The calculated magnetic moments 
($\mu_B$ per cell) for La$_2$Ba$_2$Cu$_2$O$_8$
for the 3 configurations of Ba positions. The occupations of
Ba/La are denoted as "pl-1" for plane-1, "col-2" for the
second column, etc... 
}
\vskip 5mm
\begin{center}
\begin{tabular}{|l|c|c|c|c|}
\hline
pl-1,col-1 & pl-2,col-1 & pl-1,col-2 & pl-2,col-2 & $m$ \\
\hline 
Ba & Ba & La & La & 0.00 \\
Ba & La & La & Ba & 0.09 \\
Ba & La & Ba & La & 0.20 \\
\hline
\end{tabular}
\end{center}
\end{table}
%
%
For the large supercells with 16 f.u. (112-site cells), we present results for
configurations with 4, 10 and 16 Ba sites 
distributed randomly or with a tendency of clustering on the 32 available 
La sites. We focus on the {\em clustered} configurations since they show
enhanced spontaneous magnetization.  
The calculations with 4 Ba sites correspond to the interesting doping
concentration $x=0.25$.

\begin{figure} 
\begin{center} 
\includegraphics[width=\hsize,width=8.cm]{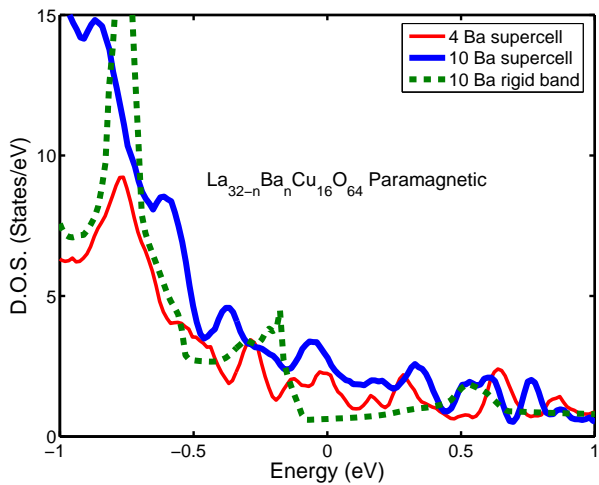} 
\end{center} 
\caption{(Color online) Total Paramagnetic DOS for the  
La$_{28}$Ba$_{4}$Cu$_{16}$O$_{64}$ 
and the La$_{22}$Ba$_{10}$Cu$_{16}$O$_{64}$ clusters
compared to a rigid band model for La$_2$CuO$_4$, in which $E_F$ is
shifted to correspond to 
$x=0.62$.} 
\label{fig1} 
\end{figure}

%
%
We first consider the
paramagnetic state for 4 and 10 clustered Ba atoms, 
where the DOS 
are calculated within a larger mesh of 125 $k$ points.
While the main DOS features with {\it random} distribution of Ba sites 
are close to a rigid band picture given by our VCA calculations and by the 
CPA results for $x=0.14$ \cite{papaconstantopoulos} and for $x=0.30$ 
\cite{bansil}, the DOS for clustered configurations can be different. The 
results for the clustered cases are compared in Fig.~\ref{fig1} together 
with DOS for the pristine phase La$_2$CuO$_4$ with a rigid band shift of 
$E_F$.
As shown in Fig.~\ref{fig1}, even a low concentration of Ba atoms is sufficient 
to induce band broadening and shifts of the DOS peaks towards $E_F$, whereas 
the rigid band model only gives a small DOS value at $E_F$. Broadening due to 
dilute substitution of La for Ba is not the whole story since the local 
DOS on individual Cu sites can also change dramatically during the 
self-consistent procedure of the spin-polarized calculations.
Curiously, the spin-polarized calculations show that the largest local 
magnetic moments and their corresponding exchange splitting of the order 
of 1 mRy do not always occur on sites with large local paramagnetic DOS. 
Therefore, while the Stoner criterion seems to provide an overall 
indicator for possible magnetization, the paramagnetic DOS may be 
different from the actual shapes of 
majority and minority DOS in the FM phase. 
These strong non-linear effects are also reflected 
by a slow convergence of the magnetic moment during the 
self-consistent procedure.
Therefore, the results for the large supercell confirm 
that the charge 
transfer from the dopants to the  
CuO$_6$ octahedra adjacent to the Ba impurities 
and the resulting spontaneous magnetization
cannot be modeled by using rigid band and 
Stoner schemes.

\begin{figure} \begin{center}
\includegraphics[width=\hsize,width=9.cm]{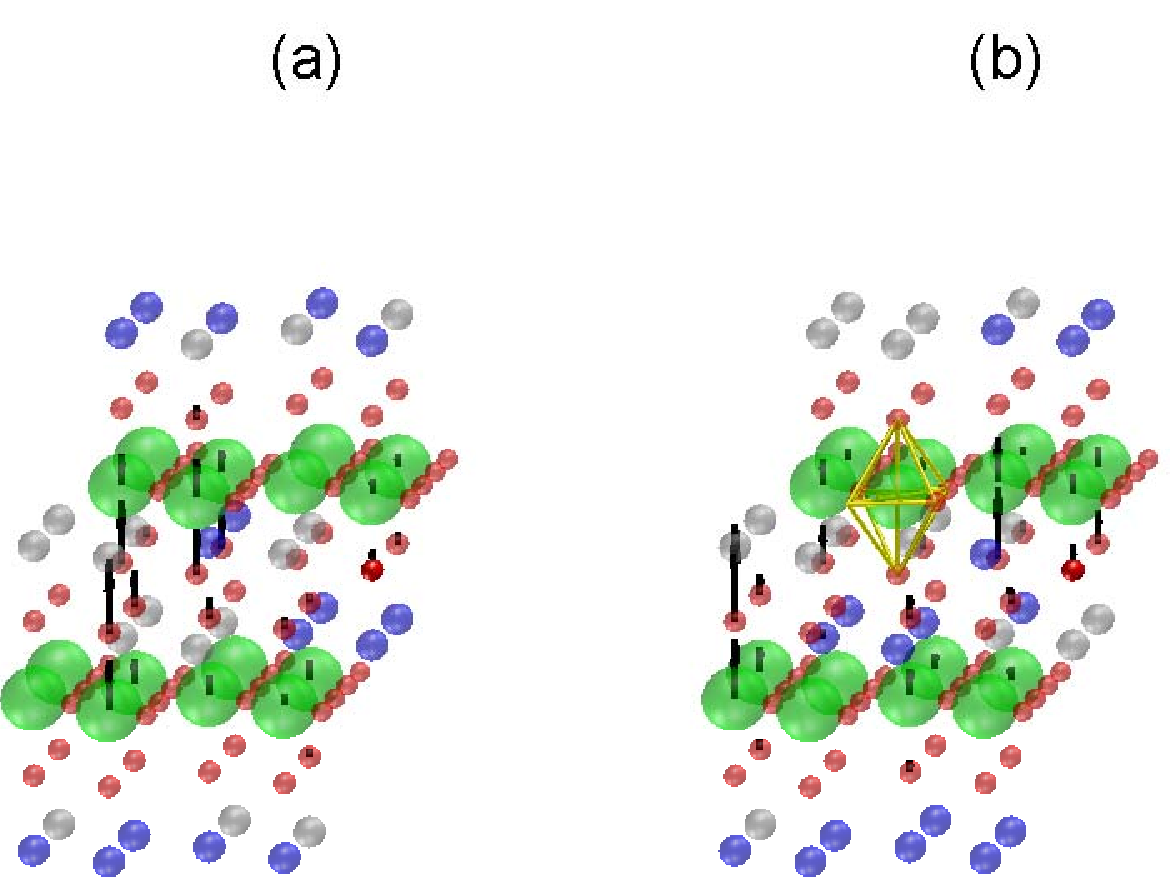}
\end{center}
\caption{
(Color online) Magnetic moments (arrows proportional
to the size of the moments) on the atoms of the  
La$_{16}$Ba$_{16}$Cu$_{16}$O$_{64}$ 
(a) random and (b) cluster 
supercells (only non-equivalent atoms 
are shown). The (gray) small spheres are Ba atoms, the small (blue) 
spheres are La atoms, the 
large (green) spheres are Cu atoms and the smallest (red) 
spheres are O atoms. The limits of an 
octahedron are outlined by yellow in (b).
The calculations have been performed with 27-$k$ points.}
\label{fig2} 
\end{figure}

\begin{figure} \begin{center}
\includegraphics[width=\hsize,width=9.cm]{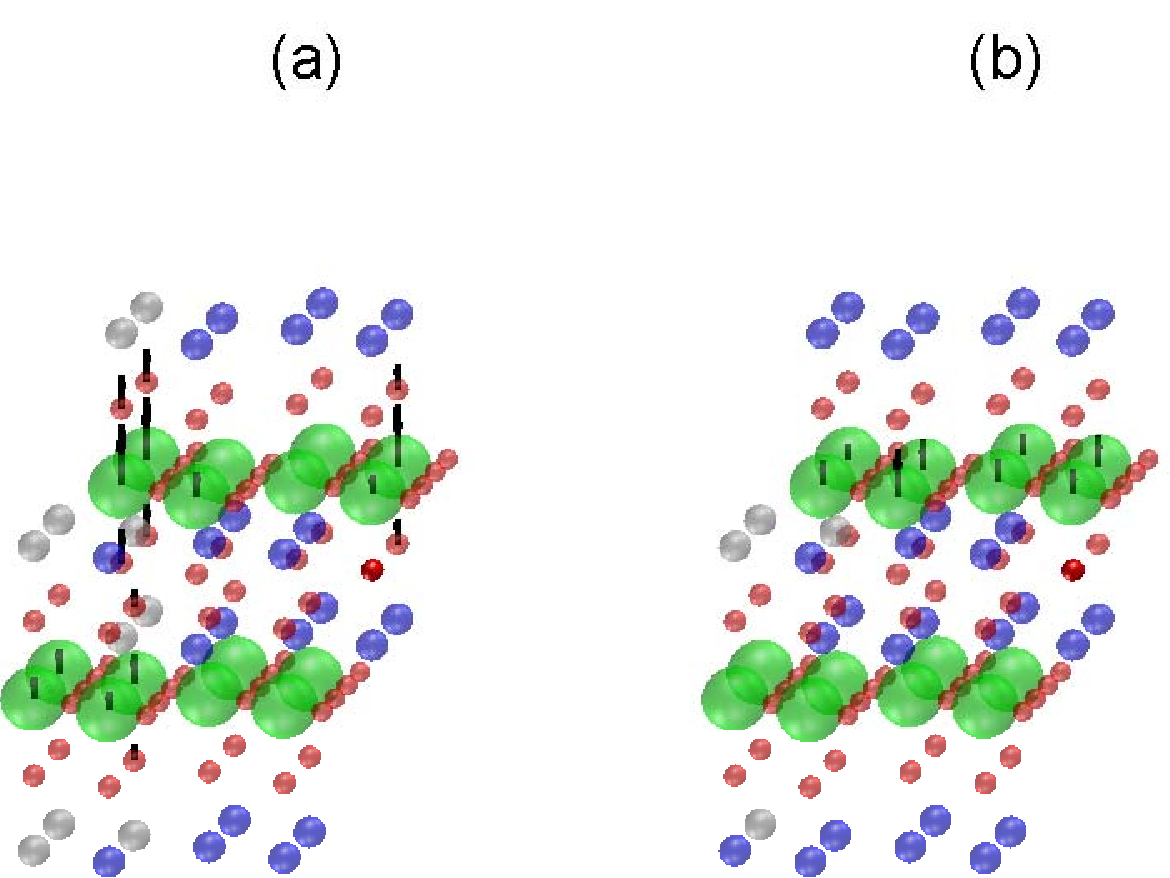}
\end{center}
\caption{(Color online) Same as Fig.~\ref{fig2} for the (a)   
La$_{22}$Ba$_{10}$Cu$_{16}$O$_{64}$ and
(b) La$_{28}$Ba$_{4}$Cu$_{16}$O$_{64}$ 
supercells.
The arrows have been rescaled by a factor 
of 2 compared to Fig.~\ref{fig2} for clarity.}
\label{fig3} \end{figure}

%
%
For the spin-polarized calculations, 
we also have studied the behavior of the magnetic 
moment as function of the number of $k$-points.
The largest magnetic moments occur for simulations 
with only 4 $k$-points. 
These calculations are analogous 
to models for
FM nano-domains \cite{kopp}, 
where the pinning 
of $E_F$ at DOS peaks 
is observed \cite{qdots}. 
However, in order to simulate bulk properties, 
we have considered computations with 12, 27 and 125
$k$-points. The total magnetic moments 
as function of $k$-points and different Ba configurations are displayed in 
Table \ref{table2}. The magnetic 
moments are localized on the Cu and apical O sites which 
are close to Ba sites, as in the case for the small cell. This can be seen 
on Fig.~\ref{fig2} and Fig.~\ref{fig3} of the supercell 
where the lengths of the arrows are proportional to the local 
magnetic moment 
on each site \cite{afm_check}.

\begin{table}[b]
\caption{\label{table2}The calculated magnetic moments 
($\mu_B$ per cell) for La$_{(2-x)}$Ba$_x$Cu$_2$O$_8$
as function of $k$-points
for 5 configurations of Ba substitution with 16, 10 and 4
Ba on the 32 possible La positions.
The averaged separation between energy levels is lower than the exchange 
splitting for the most dense $k$-point mesh, which indicates a sufficient 
$k$-convergence.}
\vskip 5mm
\begin{center}
\begin{tabular}{|l|c|c|c|c|}
\hline
$x$-configuration & 4-$k$ & 12-$k$ & 27-$k$ & 125-$k$ \\
\hline 
$x=1.0$ cluster (16 Ba) & 3.36 & 2.59  & 0.55 & 0.54\\
$x=1.0$ random (16 Ba) & 4.17 & 2.34 & 0.48 & 0.43 \\
$x=0.62$  cluster (10 Ba) & 1.05 & 0.72 & 0.24 & 0.22 \\
$x=0.25$ cluster (4 Ba) & 0.82 &  0.27 & 0.11 & 0.05\\
$x=0.25$ random (4 Ba) & 0.00 &  0.00 & 0.00 & 0.00\\ \hline
\end{tabular}
\end{center}
\end{table}

As shown in Fig.~\ref{fig2},
both the 16 Ba-site supercells, random and cluster, yield magnetic 
solutions. Therefore, a random distribution
of Ba sites does not destroy magnetic solutions for very high Ba doping. 
Fig.~\ref{fig2} confirms that the apical O sites 
close to Ba impurities have the highest magnetic moments.

Fig.~\ref{fig3} illustrates the appearance of spontaneous magnetic 
moments for lower Ba concentrations. 
When decreasing Ba concentration to 10 clustered atoms, the apical O sites do not 
develop large magnetic moments as for the configurations with 16 Ba atoms.
The effect of clustering is also seen in the case with four Ba atoms, 
i.e. for doping corresponding to the 
interesting range of doping for "overdoped" superconducting samples. 
With four clustered Ba atoms in 
the 112-site cell (3 next each other and the 4th one opposite in the next layer) 
it is possible to follow a contamination of the 
induced magnetic moment towards the Cu atoms in the second layer as shown in 
Fig.~\ref{fig3}(b). 
About 60 \% of the total magnetic moment
is on this Cu plane, which has less electrons than the first one.
Interestingly, this self-consistent 
result has been obtained from 
a starting configuration where 
the magnetic moments were put near the Ba in the first layer of Cu. 
The magnetic moments on the O atoms 
(apical and planar) are generally very small at this low doping.
This is in contrast to the result at larger doping,  where apical
sites could act as "links" between magnetic layers and thereby acquire
magnetic moments.
When the four Ba atoms are distributed randomly 
(i.e. not next to each other)
the magnetic moment goes to zero (cf. Table \ref{table2}). 
This underlines the necessity for having a rather strong, 
local perturbation (a few Ba atoms within the same plane) 
in order to polarize neighboring Cu atoms.

Our results agree with the experimental findings  
that magnetization starts near $x=0.22$ and
corroborate the interpretation of neutron scattering data
in terms of inhomogeneities such as Ba clusters 
and FM nano-domains \cite{wakimoto}. 
The observed large increase of the magnetic moment 
with doping  \cite{wakimoto} 
may indicate a substantial clustering. 
In our calculations, the highest magnetic moment 
for alternate 
La/Ba layers at $x$=1 (Table \ref{table1}) corresponds 
to $m=1.6 \mu_B$ per 
large cell instead of $m=0.43 \mu_B$ for 
the random configuration in Table \ref{table2} \cite{footnote_omag}. 
Furthermore, the case 
with $x=0.25$ and 4 $k$-points, which can simulate 
an isolated FM nano-domain as mentioned above, 
gives a magnetic moment of $m=0.82 \mu_B$ which is close 
to the experimental value $m= 0.5 \mu_B$ given by Wakimoto {\em et al.} 
\cite{wakimoto} for Sr doping that exceeds $x=0.22$.

In conclusion, we have shown that the commonly used rigid band approximation
is not appropriate for studies of doping in La$_2$CuO$_4$, at least not
for delicate magnetic properties. Our calculations with different
Ba compositions in small and large supercells show clearly different FM
solutions as function of the distribution of Ba atoms. Thus, not only
the amount of doping (Ba concentration), but also the ordering of the
Ba atoms is important. In particular, we have shown that FM can be induced
when the doping is high, and even at $x=0.25$ if there is some
clustering of the Ba atoms. In the latter case, only
some regions near the Ba clusters are weakly polarized.
The mechanism of FM from Ba- (or Sr-) clusters could be responsible 
for the suppression of superconductivity in "overdoped" cuprates 
as suggested by Kopp {\em et al.}\cite{kopp}.

We thank useful discussions with R.S. Markiewicz and
C. Berthod. This work was supported by the US Department 
of Energy contracts DE-AC03-76SF00098 and DE-FG02-07ER46352 and
benefited from the allocation of supercomputer 
time at the NERSC and the Northeastern University's Advanced Scientific 
Computation Center (NU-ASCC).


\begin{thebibliography}{99}
\bibitem{footnote1} The highest $\ell$ is obtained via tails and 
3 centre terms.
\bibitem{afm_check} We have checked that 
small negative moments on Cu sites do not grow 
during the iterations. We therefore discard AFM 
solutions at these Ba concentrations.
\bibitem{footnote_omag}
Hybridization between the apical oxygen 2$p_z$ state and Cu $d$ 
states of magnetic Cu sites contributes to the 
magnetization on some of the apical oxygen. 
However, pair-breaking within the superconducting density 
is due to the magnetism within the Cu planes.
\end{thebibliography}
\end{document}